\documentclass[twocolumn,pre,preprintnumbers]{revtex4}
\usepackage{textcomp}
\usepackage{amsmath}
\usepackage{amssymb}
\usepackage{graphicx}
\usepackage{color}
\usepackage{soul}
\newcommand{\be}{\begin{equation}}
\newcommand{\ee}{\end{equation}}
\newcommand{\ba}{\begin{eqnarray}}
\newcommand{\ea}{\end{eqnarray}}

\begin{document}

\title{Classical-quantum crossover in the critical behavior of the transverse field S-K spin glass model}

 \author{Sudip Mukherjee}
 \email{sudip.mukherjee@saha.ac.in}
 \author{Atanu Rajak}
 \author{Bikas K Chakrabarti}
 \affiliation{Condensed Matter Physics Division, Saha Institute of Nuclear Physics, 1/AF Bidhannagar, Kolkata 700064, India }

 \begin{abstract}
 We study the critical behavior of Sherrington-Kirkpatrick model in transverse field (at finite temperature)  using Monte Carlo simulation and exact diagonalization (at zero temperature). We determine the phase diagram of the model by estimating the Binder cumulant. We also determine the correlation length exponent  from the collapse of the scaled data. Our numerical studies here indicate that critical Binder cumulant (indicating the universality class of the transition behavior) and the correlation length exponent cross over from their `classical' to  `quantum' values at a finite temperature (unlike the cases of pure systems where such crossovers occur at zero temperature). We propose a qualitative argument supporting such an observation, employing a simple tunneling picture.

\end{abstract}
\maketitle

\section{Introduction}
The motivation of this work is to study the phase diagram and critical behavior of the Sherrington-Kirkpatrick (S-K) spin glass model~\cite{sudip-sk} in transverse field~\cite{sudip-bkc-book} using Monte Carlo and exact diagonalization techniques at finite and zero temperature respectively and investigate the crossover behavior from classical to quantum fluctuation dominated phase transitions. Several approximate theoretical and numerical studies (see Refs.~\cite{sudip-yamamoto,sudip-usadel,sudip-kopec,sudip-gold,sudip-lai}) have already been made on S-K model to get some isolated features of the quantum phase transition of this model. We report here a detailed numerical study. Using both Monte Carlo and exact  diagonalization we determine the critical Binder cumulant~\cite{sudip-binder} which is an indicator of the nature of critical fluctuation. It also provides critical transverse field or temperature. We study the scaling behavior of the Binder cumulants with respect to the system sizes 
and the scaling fit gives the value(s) of the correlation length exponent.  We find 
critical Binder cumulant 
and 
correlation 
length exponent cross over from a `classical' value (corresponding to the classical S-K model) for high temperature and low transverse field, to a `quantum' value for low temperature and high transverse field at a finite temperature.
\section{Model}
\label{model}

The Hamiltonian of quantum S-K model of $N$ spins is given by
\begin{align}
H  = H_0 + H_I;~ H_0=-\sum_{\langle i,j\rangle} J_{ij}\sigma_i^z\sigma_j^z;~ H_I=-{\Gamma} \sum_{i = 1}^N\sigma_i^x , \label{Ham}  
\end{align}
where $\sigma_i^z$, $\sigma_i^x$ are the $z$ and $x$ components of Pauli spin matrices respectively and $\Gamma$ is the transverse field. For $\Gamma=0$ the Hamiltonian in Eq.~(\ref{Ham}) reduces to the classical S-K spin glass Hamiltonian ($H_0$). In this model spin-spin interactions ($J_{ij}$) are distributed  following Gaussian distribution $\rho (J_{ij}) = \Big (\frac{N}{2{\pi}J^2}\Big)^{\frac{1}{2}}\exp\Big (\frac{-NJ_{ij}^2}{2J}\Big)$. The mean of Gaussian distribution is zero and the variance is $J/ \sqrt{N}$. We work with $J = 1$. The effective classical Hamiltonian $H_{eff}$ of the Hamiltonian in Eq.~(\ref{Ham}) can be obtained by using Suzuki-Trotter formalism~(see e.g., \cite{sudip-bkc-book}):
\begin{align}
 H_{eff}=-\sum_{n=1}^M \sum_{\langle i,j\rangle} {J_{ij}\over M}\sigma_i^n\sigma_j^n-\sum_{i=1}^N\sum_{n=1}^M{1\over {2\beta}}\text{log~coth}{\beta\Gamma\over M}\sigma_i^n\sigma_i^{n+1} \label{H_cl},
\end{align}
where $\sigma_i^n=\pm 1$ is the classical Ising spin and $\beta$ is the inverse of temperature $T$. The additional dimension appears in Eq.~\ref{H_cl}, is often called Trotter direction (theoretically ${M\to \infty}$).

\begin{figure*}
 \begin{center}
 \includegraphics[width=5.5cm]{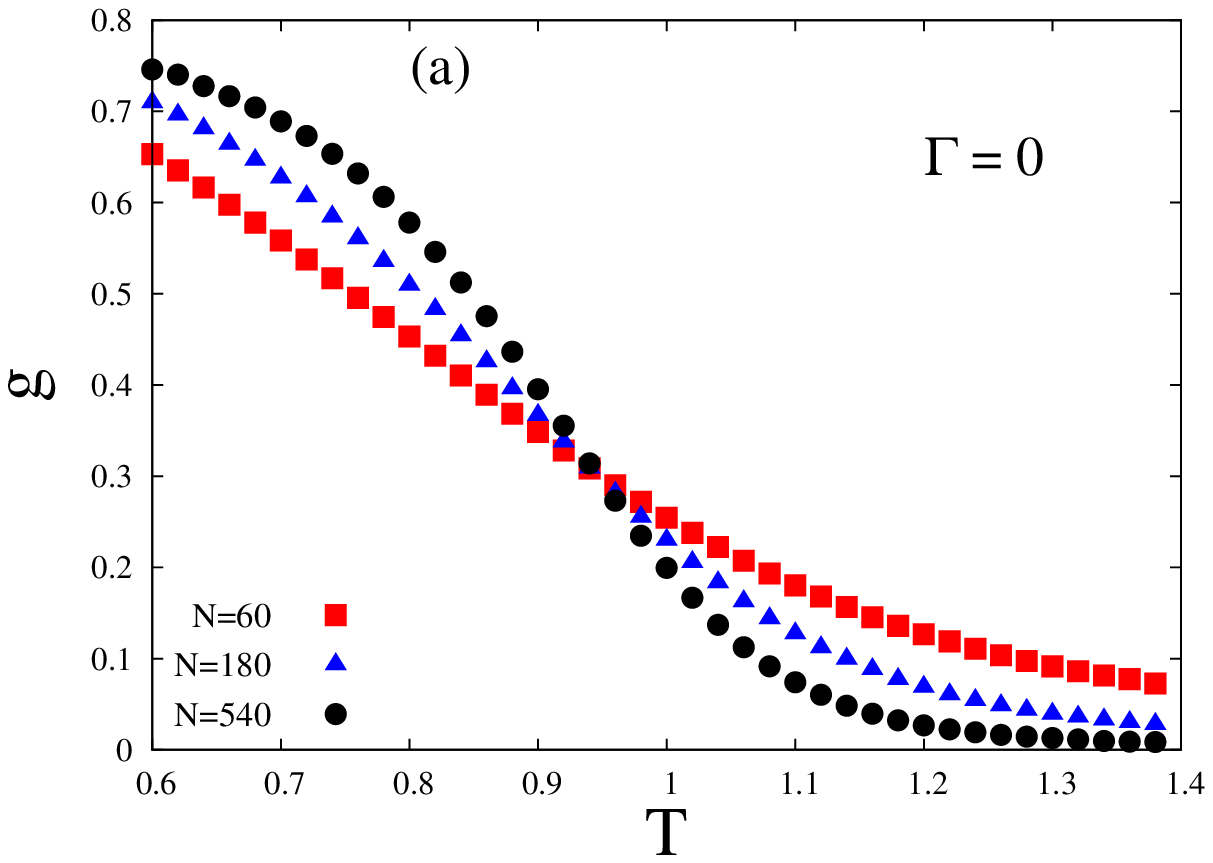}
 \includegraphics[width=5.5cm]{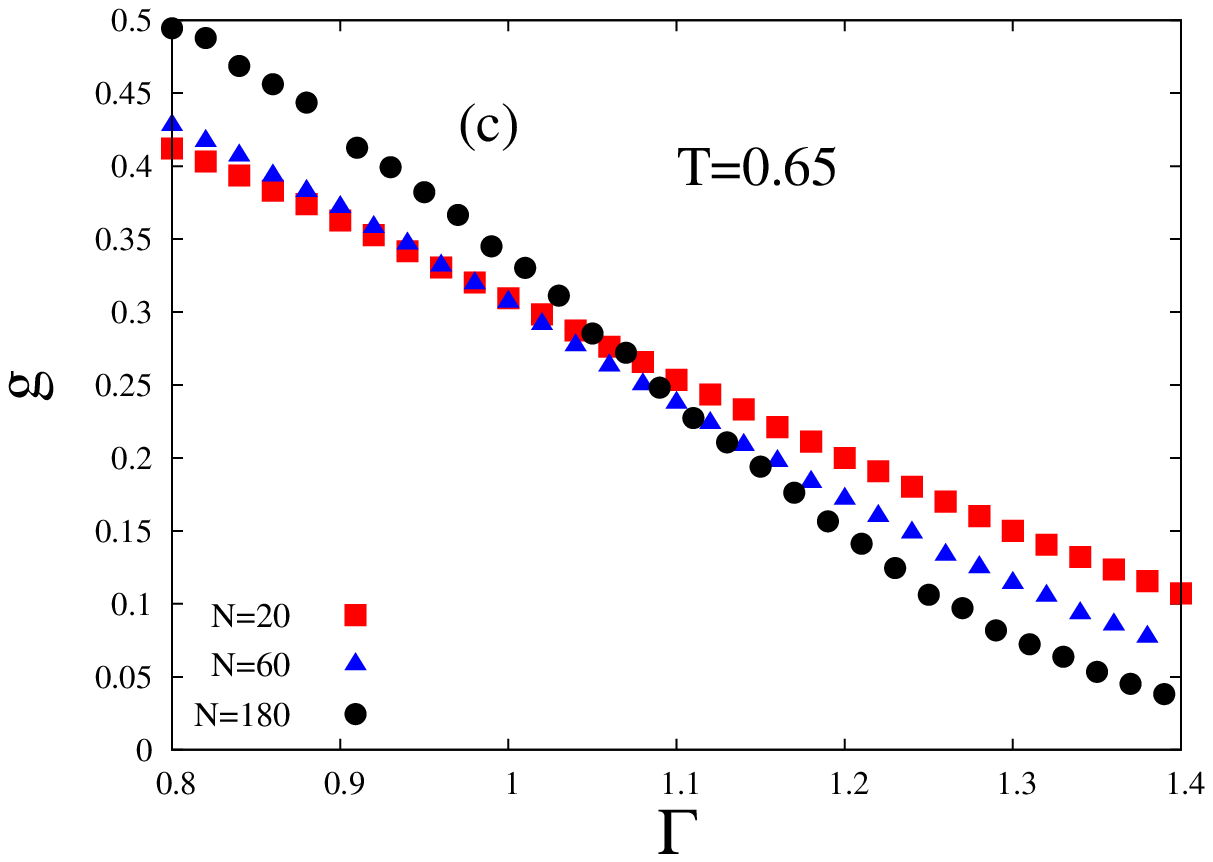}
 \includegraphics[width=5.5cm]{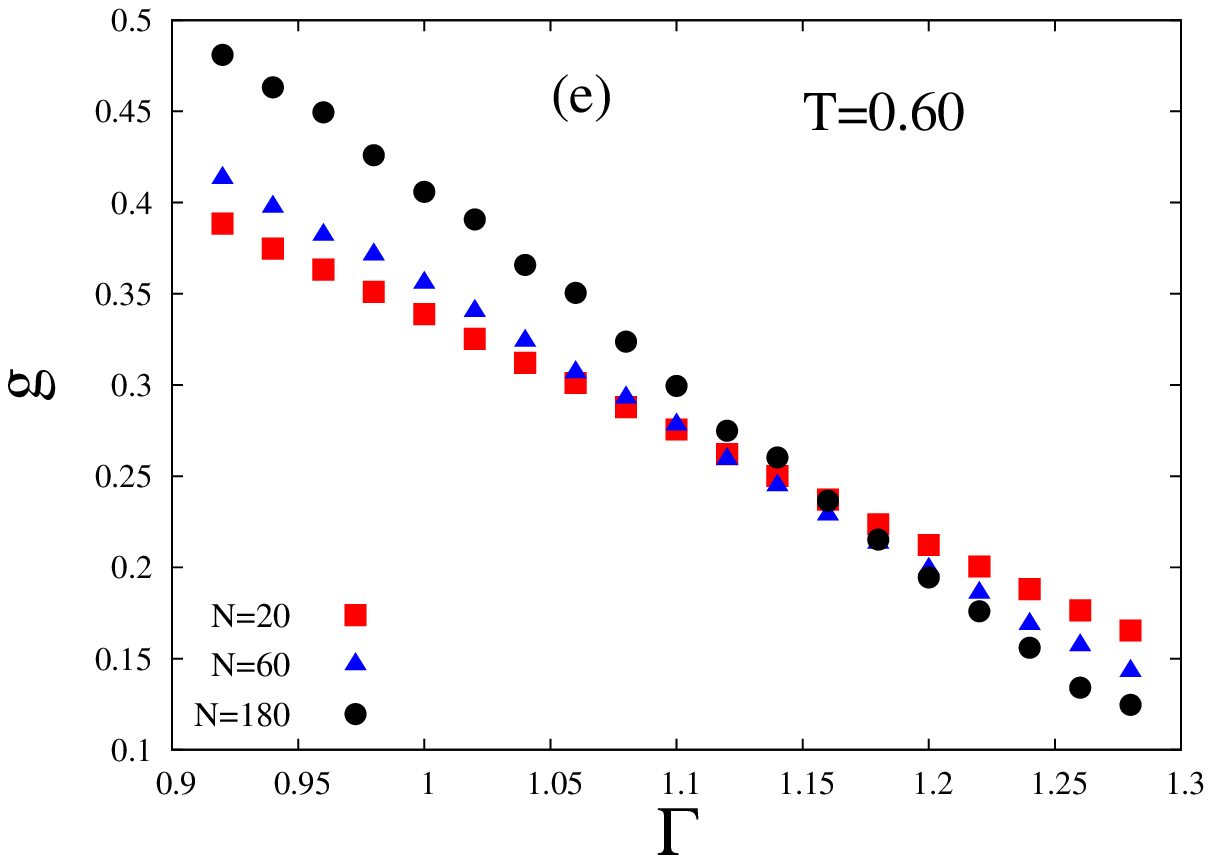}
 \includegraphics[width=5.5cm]{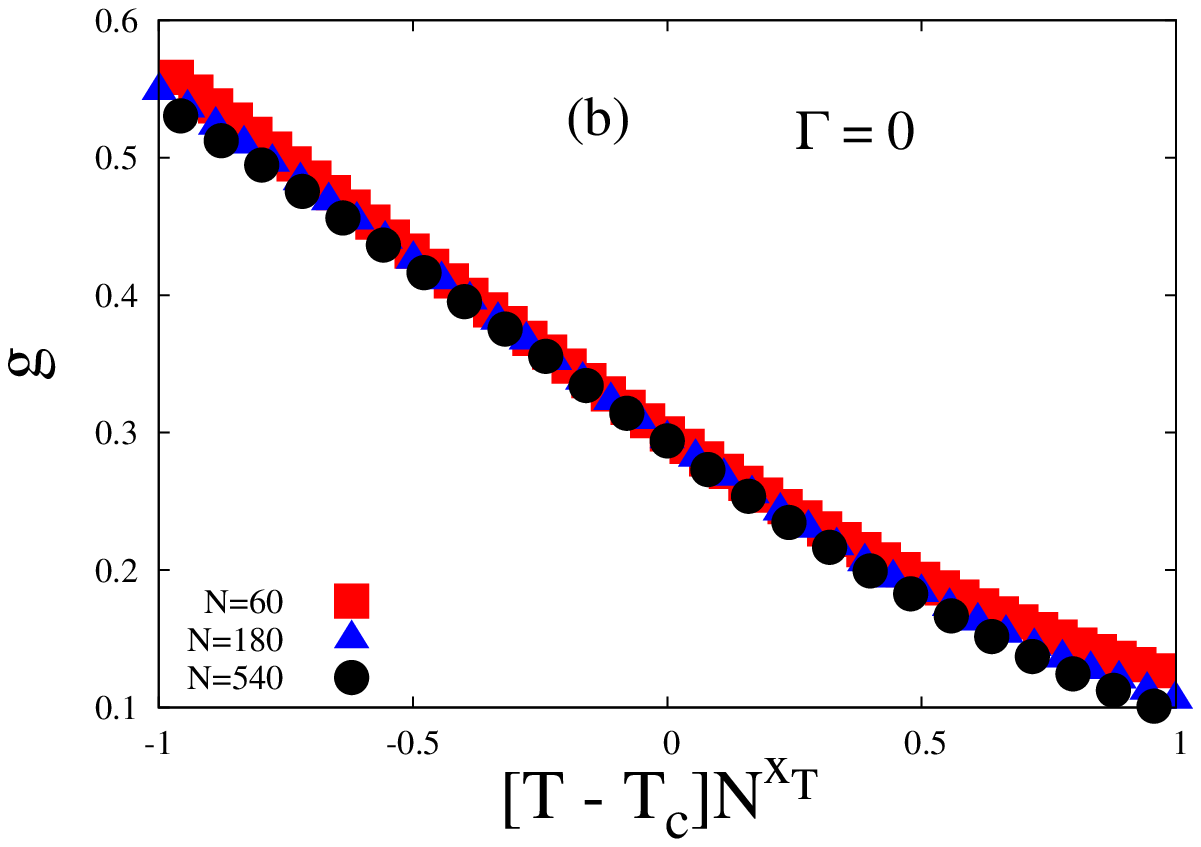}
 \includegraphics[width=5.5cm]{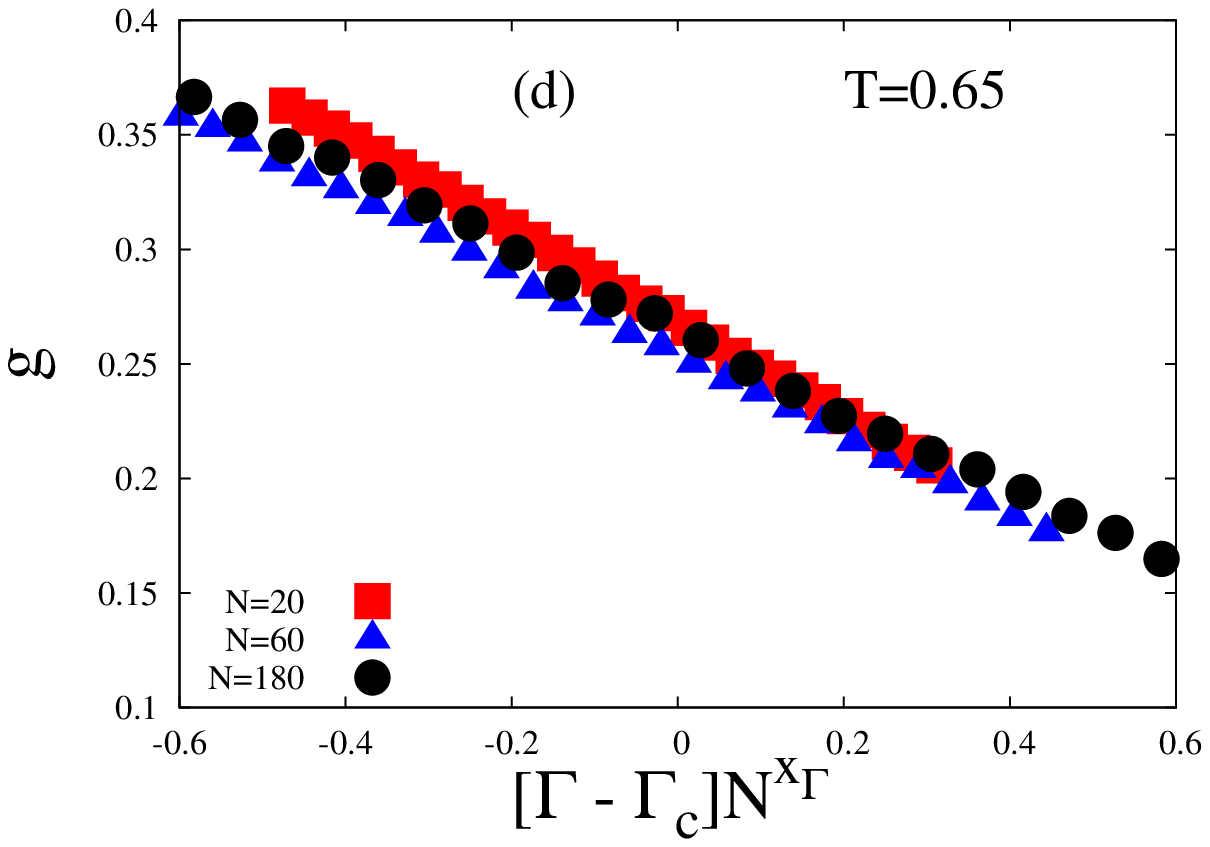}
 \includegraphics[width=5.5cm]{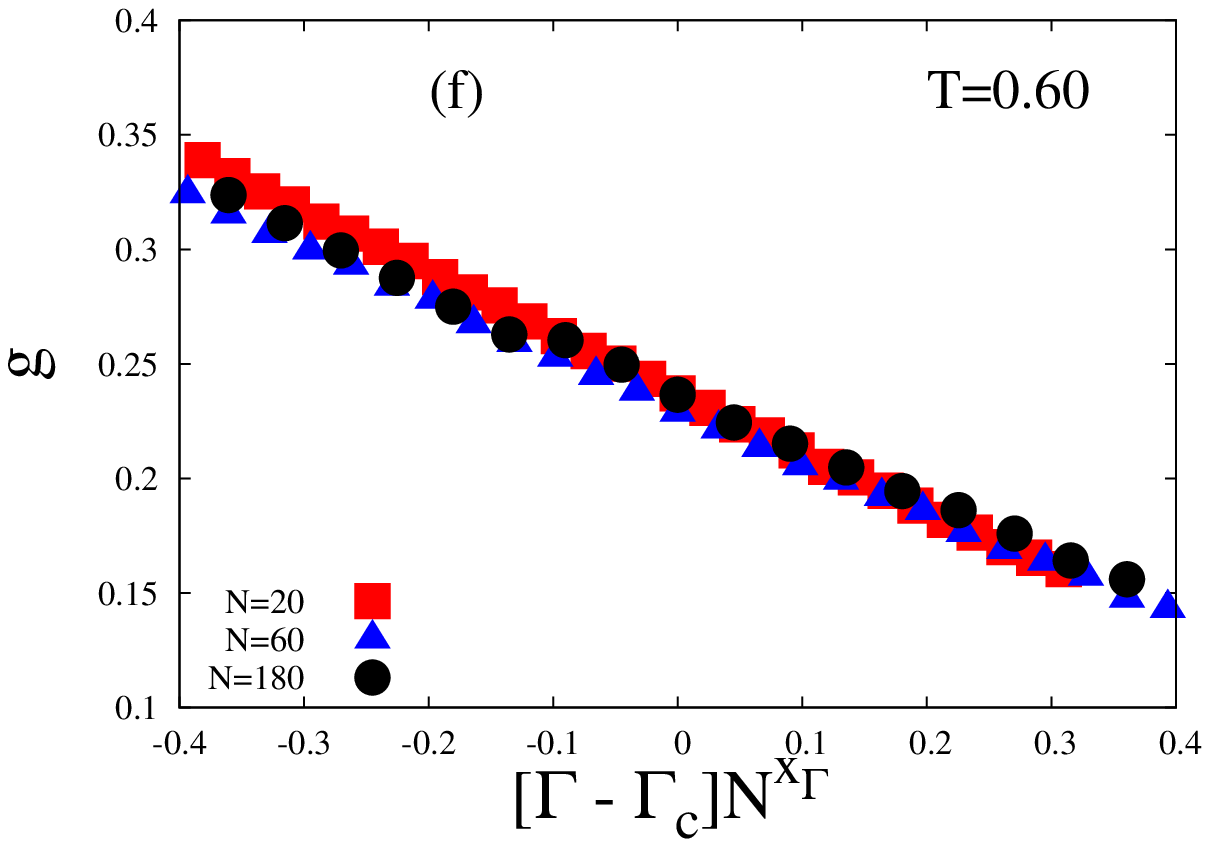}
 \end{center}
 \caption{
Monte Carlo results for the Binder cumulant ($g$) plotted as  function of temperature $T$ and transverse field $\Gamma$ are shown: (a) for classical S-K model (at $\Gamma=0$) and (c) and (e) for $T = 0.65$ and $0.60$ respectively.  The crossing points give the estimate for $T_c$ or $\Gamma_c$. The statistical errors are indicated by the symbol sizes. Figs.~(b), (d) and (f) show the  collapses of $g$ curves of (a), (c) and (e) respectively when the variations of $g$ are plotted against $[T - T_c]N^{x_T}$ or $[\Gamma - \Gamma_c]N^{x_{\Gamma}}$ (see Eq.~\ref{gt}). The scaling collapses  give the  values  $x_T$ or $x_{\Gamma}$ $=0.31 \pm 0.02$.
 }
 \label{inset_high1}
 \end{figure*}

\begin{figure*}
\begin{center}
\includegraphics[width=5.7cm]{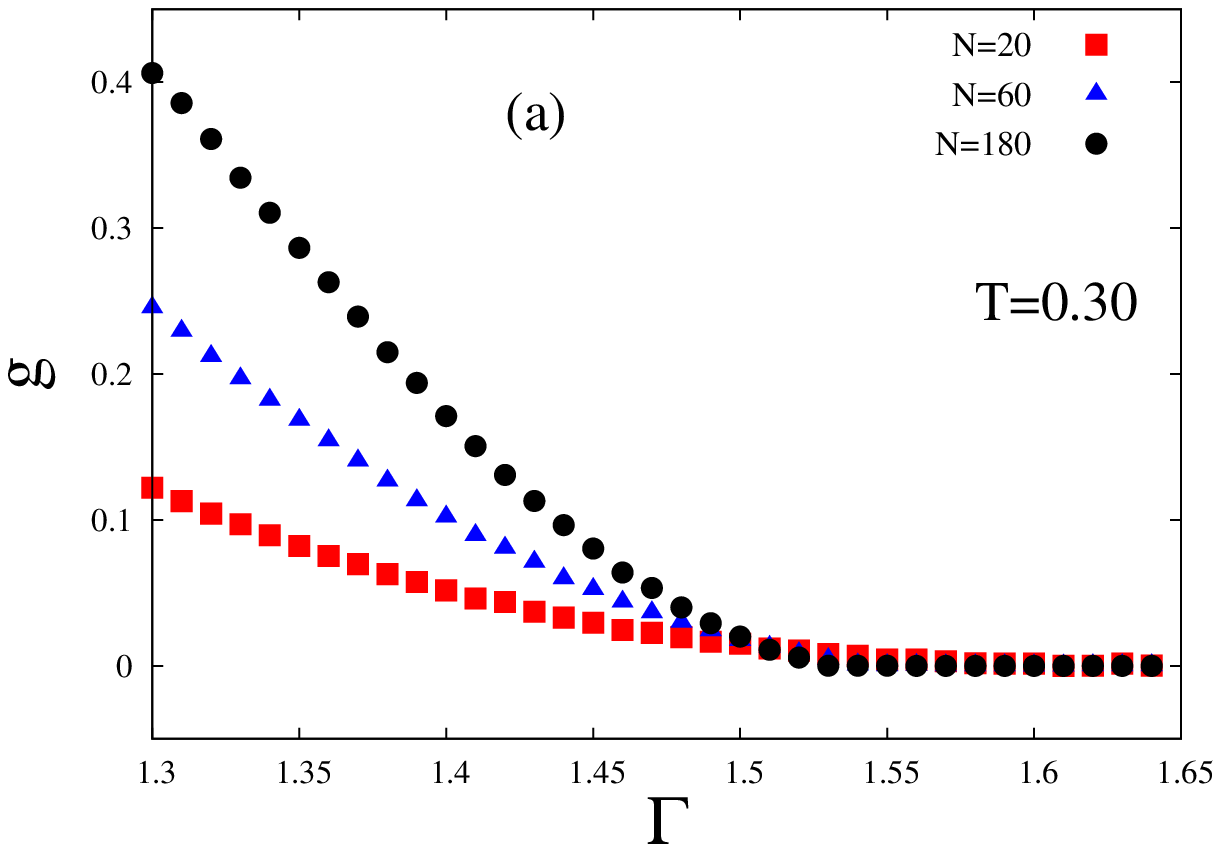}
\includegraphics[width=5.7cm]{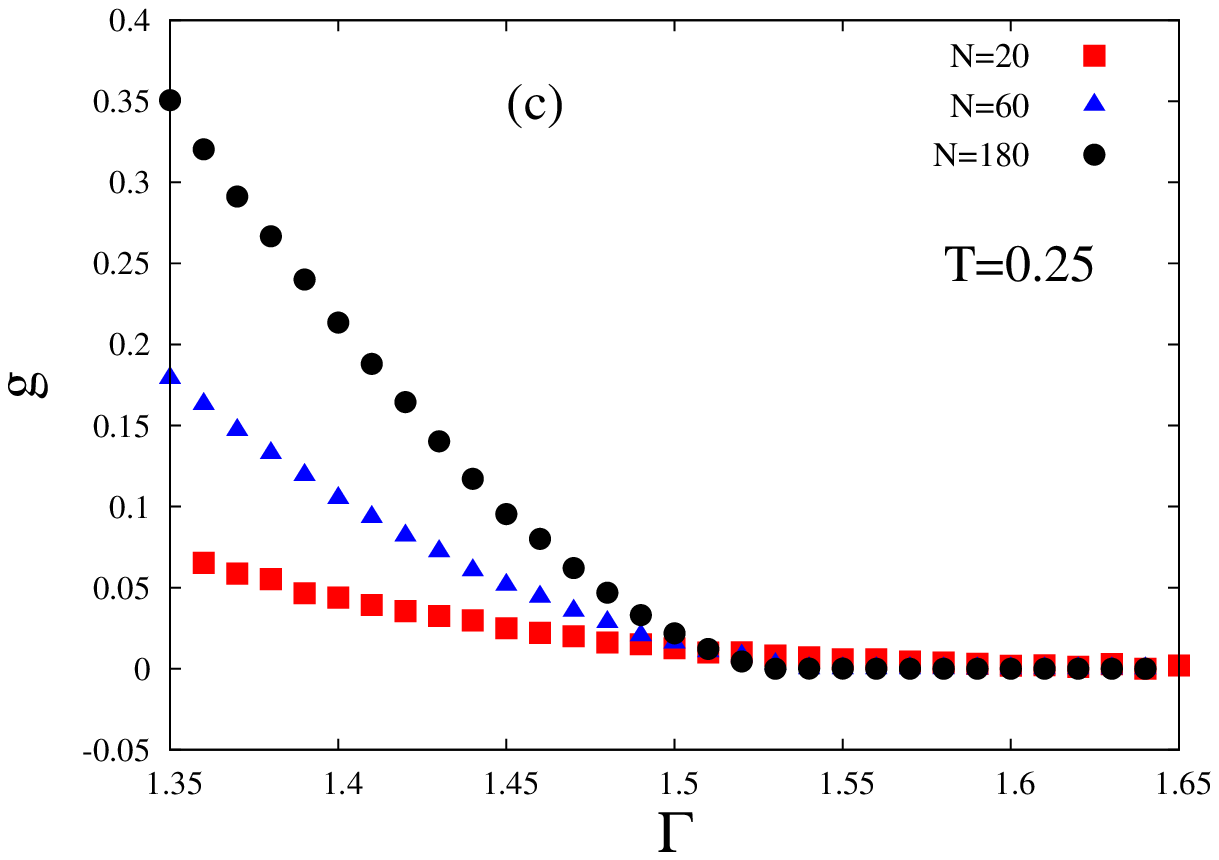}
\vskip 0.1cm
\includegraphics[width=5.7cm]{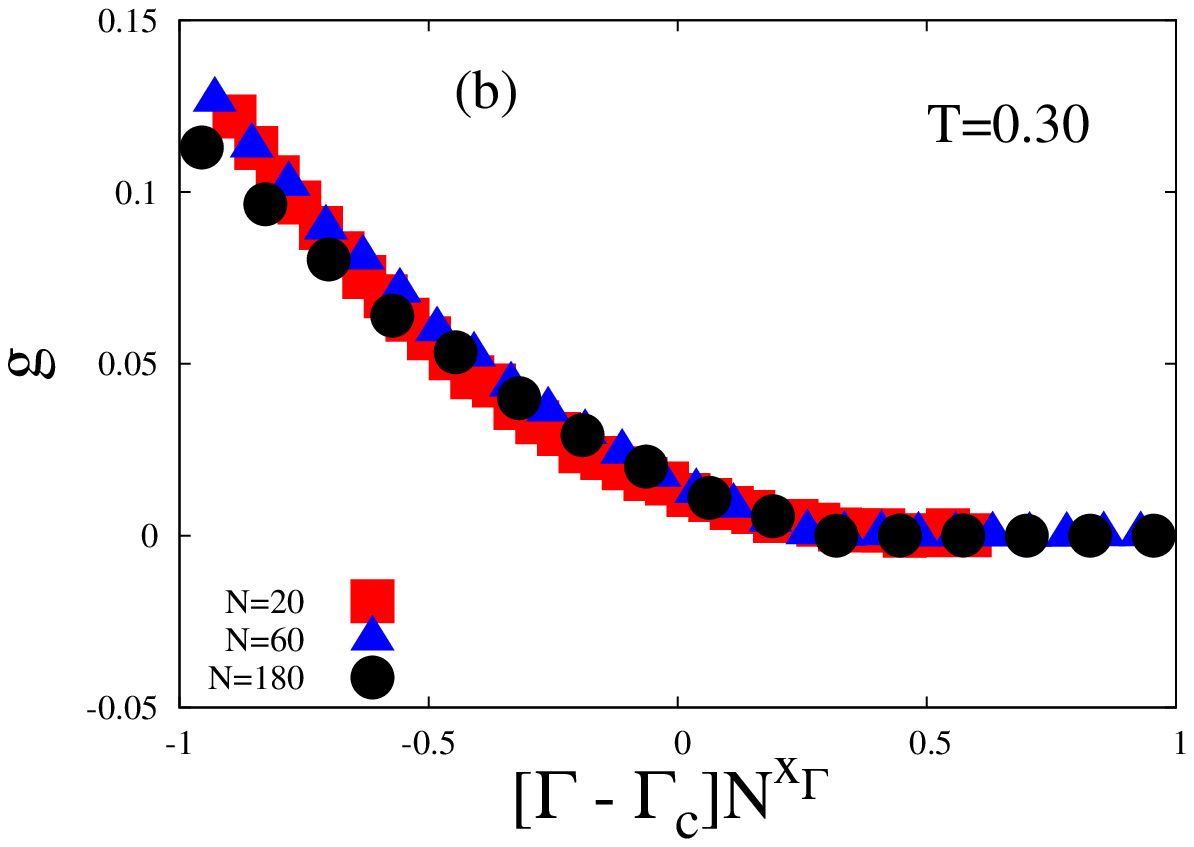}
\includegraphics[width=5.7cm]{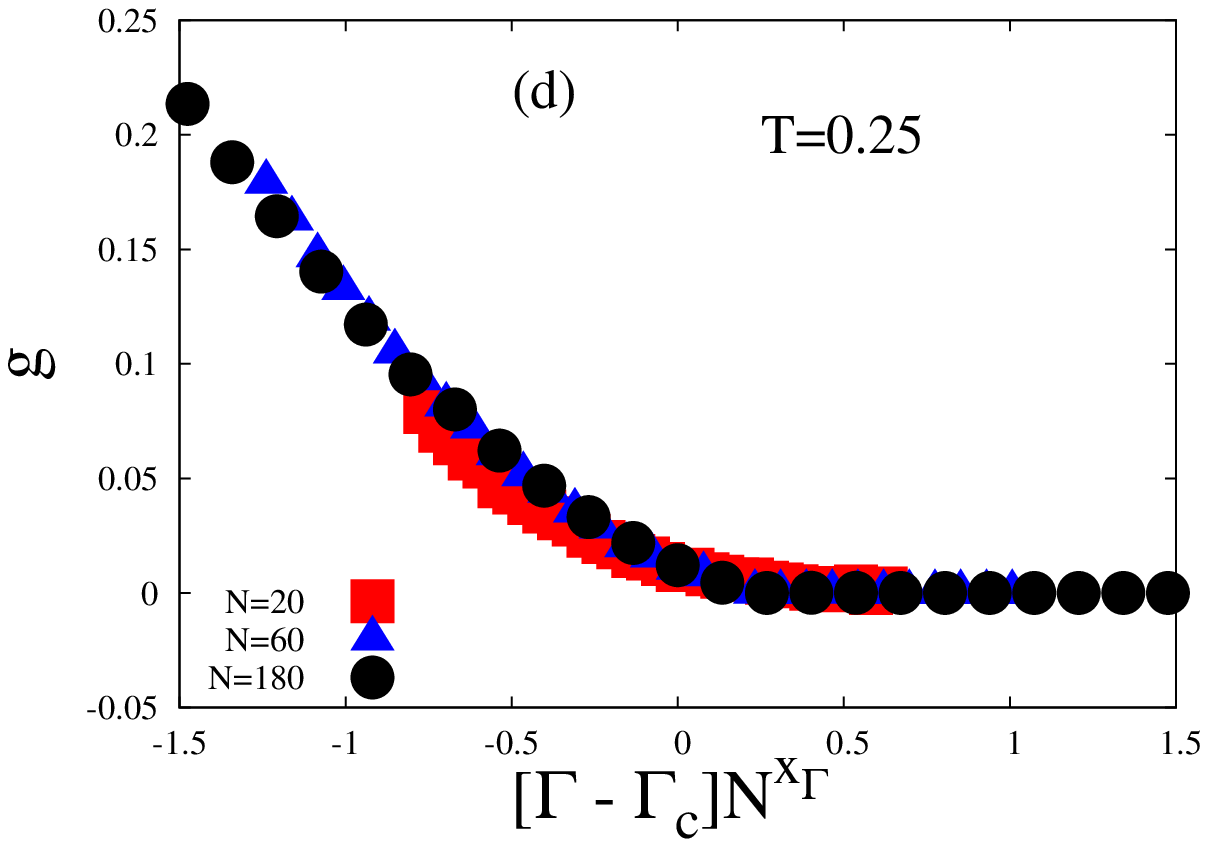}
\end{center}
\caption{ 
Monte Carlo results for the Binder cumulant ($g$) plots with transverse field $\Gamma$ at temperatures $0.30$ and $0.25$  are shown in (a) and (c) respectively. The statistical errors are of the order of the symbol sizes. 
Figs.~(b) and (d)  show the collapses of $g$ curves in (a) and (c)  respectively. Again the 
variations of $g$ are plotted according to the scaling relation Eq.~(\ref{gt}) and the collapses  give the  value   $x_{\Gamma}$ $=0.50 \pm 0.02$.
}
 
\label{inset_low1}
\end{figure*}


\section{Monte Carlo Results}
We accomplish Monte Carlo simulation using Hamiltonian in Eq.~(\ref{H_cl}) to find the critical transverse field for a fixed temperature. We also perform Monte Carlo simulation on Hamiltonian $H_0$ to extract the critical behavior of the classical S-K model. We take  $t_0$ Monte Carlo steps to equilibrate the system and  make  Monte Carlo averaging over next $t_1$  steps. To study the critical behavior of the model, we take replica overlap $q$, which is defined as $q=\frac{1}{NM}\sum_{i=1}^N\sum_{n=1}^M(\sigma_i^n(t))^{\phi}(\sigma_i^n(t))^{\theta}$, where $(\sigma_i^n)^{\phi}$ and $(\sigma_i^n)^{\theta}$ are the spins of two different replicas $\phi$ and $\theta$ corresponding to the  same realization of disorder.
We study the variation of average Binder cumulant ($g$) with $\Gamma$ and $T$ for different system sizes. For our study we define the average Binder cumulant~\cite{sudip-guo,sudip-alvarez} given by:
\begin{align}
 g=\frac{1}{2}\Big[3-\overline{\Big(\frac{\langle q^4 \rangle}{(\langle q^2 \rangle)^2}\Big)}\Big] ,\label{g2}
\end{align}
where $\langle.\rangle$ and overhead bar indicate thermal and configuration averages respectively.
It may be noted that with another definition for disorder averaging \cite{sudip-guo} $ g=\frac{1}{2}\Big[3-\frac{\overline{\langle q^4 \rangle}}{\overline{(\langle q^2 \rangle)^2}}\Big]$ one obtains huge fluctuation and bad statistics (see e.g.,~\cite{sudip-guo}).
 We therefore work with the above definition of $g$ (Eq.~(\ref{g2})) to make a consistent study through out the entire range of temperature.
 
 Near critical point $g$ scales as  $g=g(L/\xi, M/L^{z})$ where $L$ denotes the linear size of the system and $M$ is the Trotter size. The dynamical exponent is symbolized by $z$.
    $\xi$ represents the correlation length, which scales as $\xi$ $\sim (T - T_c)^{-\nu_{T}}$ or 
 $(\Gamma - \Gamma_c)^{-\nu_{\Gamma}}$ with correlation exponents 
$\nu_{T}$ or $\nu_{\Gamma}$. Hence close to critical region we can write, 
  \begin{equation}
g \sim g((T - T_c)N^{x_T},M/N^{z/d_c})~\text{or}~g((\Gamma - \Gamma_c)N^{x_{\Gamma}},M/N^{z/d_c}) \label{gt}
\end{equation}
where $x_T= 1/\nu_Td_c$ and $x_{\Gamma}= 1/\nu_{\Gamma}d_c$ with $L=N^{1/d_c}$. The critical transverse field or temperature are denoted by $\Gamma_c$ or $T_c$ respectively and $d_c$ is the effective dimension of the system.
The intersection of the $g$ vs. $\Gamma$ curves for different system sizes (keeping $M/L^{z}$ fixed) gives the estimate of  values of  
  $\Gamma_c$ and critical Binder cumulant  $g_c$.  We try to collapse the $g$ curves by following the Eq.~(\ref{gt}). Such collapses of the $g$ curves are made by suitably scaling the tuning parameters with chosen values of the exponents $x_T$ and $x_{\Gamma}$. 
 
 To simulate $H_{eff}$ we take system sizes $N=20, 60, 180$. We work with $d_c=6$ and $z=4$~\cite{sudip-Billoire} (these values are associated with classical S-K model). To keep $M/L^{z}$ fixed, we start with $M=10$ for the system size $N=20$ and take $M=21, 43$ for system sizes $N=60, 180$ respectively. Due to the absence of any additional dimension (Trotter dimension) in the Hamiltonian $H_0$, we are able to  take larger system sizes $N=60, 180, 540$ in the Monte Carlo simulation of the classical S-K model. The equilibrium time of the system is
 $t_0=75000$ and we take $25000$ ($t_1$) Monte Carlo steps  for thermal averaging. $1000$ samples are averaged over to get the configuration 
 average. We notice that in the range starting from the classical S-K model at $\Gamma=0$ to almost $T\simeq0.50$ ($\Gamma \simeq 1.30$),  the $g_c$  takes a constant value $0.22 \pm 0.02$ (see  Fig.~\ref{inset_high1} (a, c, e)) and  we  find good  data collapse of g curves (to Eq.~\ref{gt}) for $x_T=x_{\Gamma}= 0.31\pm 0.02$ (see Fig.~\ref{inset_high1} (b, d, f)). This result ($x_T=x_{\Gamma}$ or $\nu_T=\nu_{\Gamma}$) is also consistent with
 analytic nature of the $T$-$\Gamma$ phase boundary of the model (see Sec.~\ref{summary}).  In the range $T=0.30$ ($\Gamma \simeq 1.50$) to $T = 0.20$ ($\Gamma \simeq 1.54$), we observe that the  value of $g_c$ is nearly equal to zero but in this range we do not get  decent collapses of $g$ curves for any one chosen value of $x_{\Gamma}$. We repeat our simulation in this range with $d_c=8$ and $z=2$~\cite{sudip-david,sudip-read} (these values correspond to quantum S-K model). With these values of $d_c$ and $z$ we take Trotter sizes $M=10, 13, 17$ for the system sizes $N=20, 60, 180$ respectively to keep $M/L^{z}$ constant. Again we find vanishingly small value of $g_c$ (see Fig.~\ref{inset_low1} (a, c)). This time we get  good 
data collapses of $g$ curves (see Fig.~\ref{inset_low1} (b, d)) for $x_{\Gamma} = 0.50 \pm 0.02$. Such a crossover in $g_c$ or the exponent value $x_{\Gamma}$ ($=x_T$) with $\Gamma$ (or $T$) values 
within this range ($0.5<T<0.35$, $1.30<\Gamma<1.45$) may be abrupt. Our numerical analysis is not very accurate here and 
gradual changes within this range cannot be ruled out.

\section{Zero-temperature diagonalization results}\label{pq}
We explore the pure quantum critical behavior of the spin glass (i.e., the system at temperature $T=0$) through the Binder cumulant 
analysis of the system using an exact diagonalization technique. We have performed exact diagonalization of the Hamiltonian for rather small 
system sizes (up to $N=22$) using Lanczos algorithm~\cite{lanczos_52}. 
Here, we are interested to show the continuity of our Monte Carlo result of nearly zero value of critical Binder cumulant even at zero temperature. 
We construct the Hamiltonian of Eq.~(\ref{Ham}) in spin basis states i.e., the eigenstates of the spin operators ($\sigma_i^{z}$, $i=1,..,N$) 
for performing the diagonalization.
Then the $n$-th eigenstate of the Hamiltonian in Eq.~(\ref{Ham}) is represented as $|\psi_n\rangle~= \sum_{\alpha=0}^{2^{N-1}} a_{\alpha}^n |\varphi_\alpha\rangle$,
where $|\varphi_\alpha\rangle$ are the eigenstates of the Hamiltonian $H_0$ and $a_{\alpha}^n=\langle\varphi_{\alpha}|\psi_n\rangle$. As we are interested in the zero temperature analysis, our main focus is confined on the ground state ($|\psi_0\rangle$) averaging of different quantities. In this case the order parameter of the system can be defined as $Q = (1/N) \sum_i \overline{\langle\psi_0|\sigma_i^z|\psi_0\rangle^2}$. 
The configuration average is again indicated by the overhead bar. To calculate Binder cumulant, the various moments can be calculated using Refs.~\cite{sudip-sk,sen_97}],
\be
 Q_k = {1\over {N^k}} \sum^{N}_{i_1} \ldots \sum^{N}_{i_k}\langle\psi_0|\sigma_{i_1}^z \ldots \sigma_{i_k}^z|\psi_0\rangle^2.
\label{moment}
\ee
Here $Q_k$s are actually $k$-spin correlation functions for a particular disorder configuration. One can easily realize that order parameter $Q=\overline{Q_1}$.
 If we know the ground state at different parameter values of the Hamiltonian the various moments can be determined using Eq.~(\ref{moment}).
In this context the average Binder cumulant is defined as $g=\frac{1}{2}\Big[3-\overline{\Big(\frac{ Q_4 }{( Q_2)^2}\Big)}\Big]$ (note the difference with the Eq.~(\ref{g2})).

\begin{figure}[ht]
\begin{center}
\includegraphics[width=5.5cm]{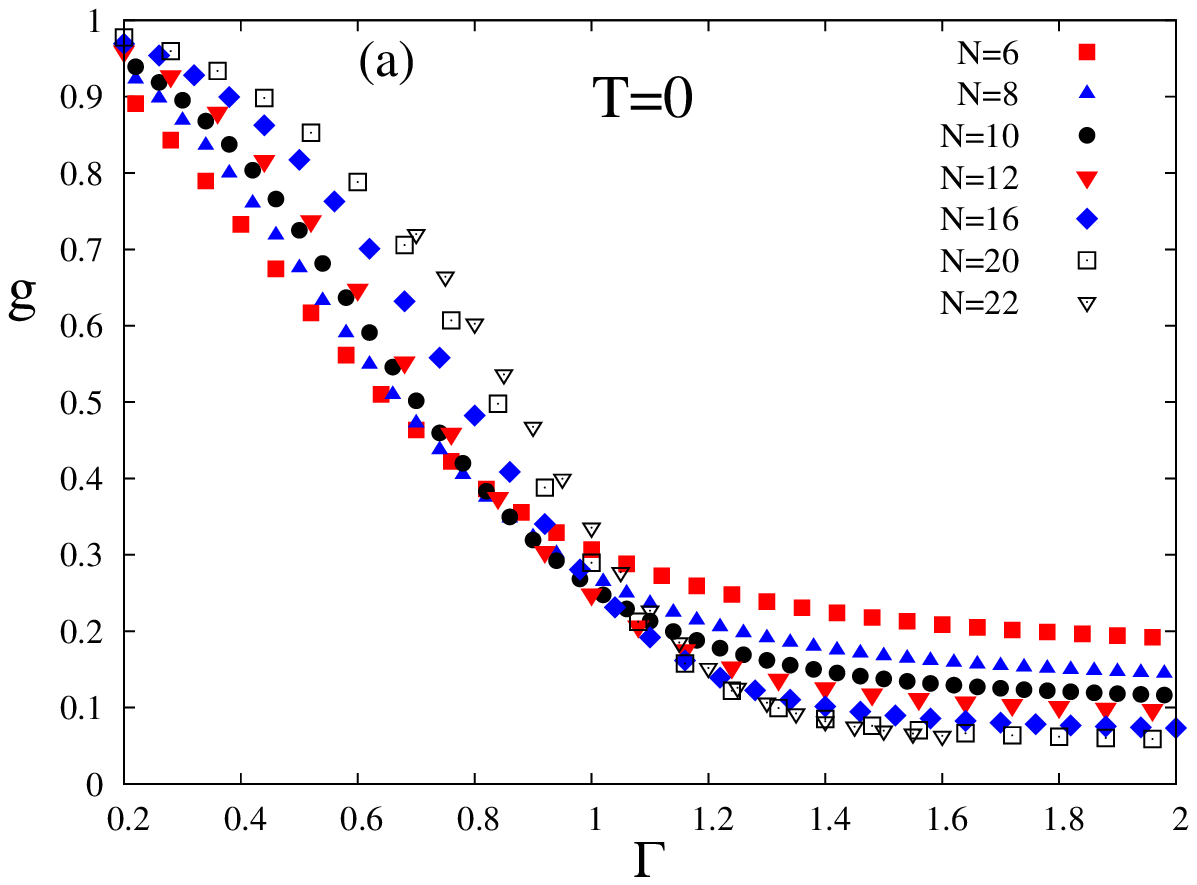}
\includegraphics[width=5.5cm]{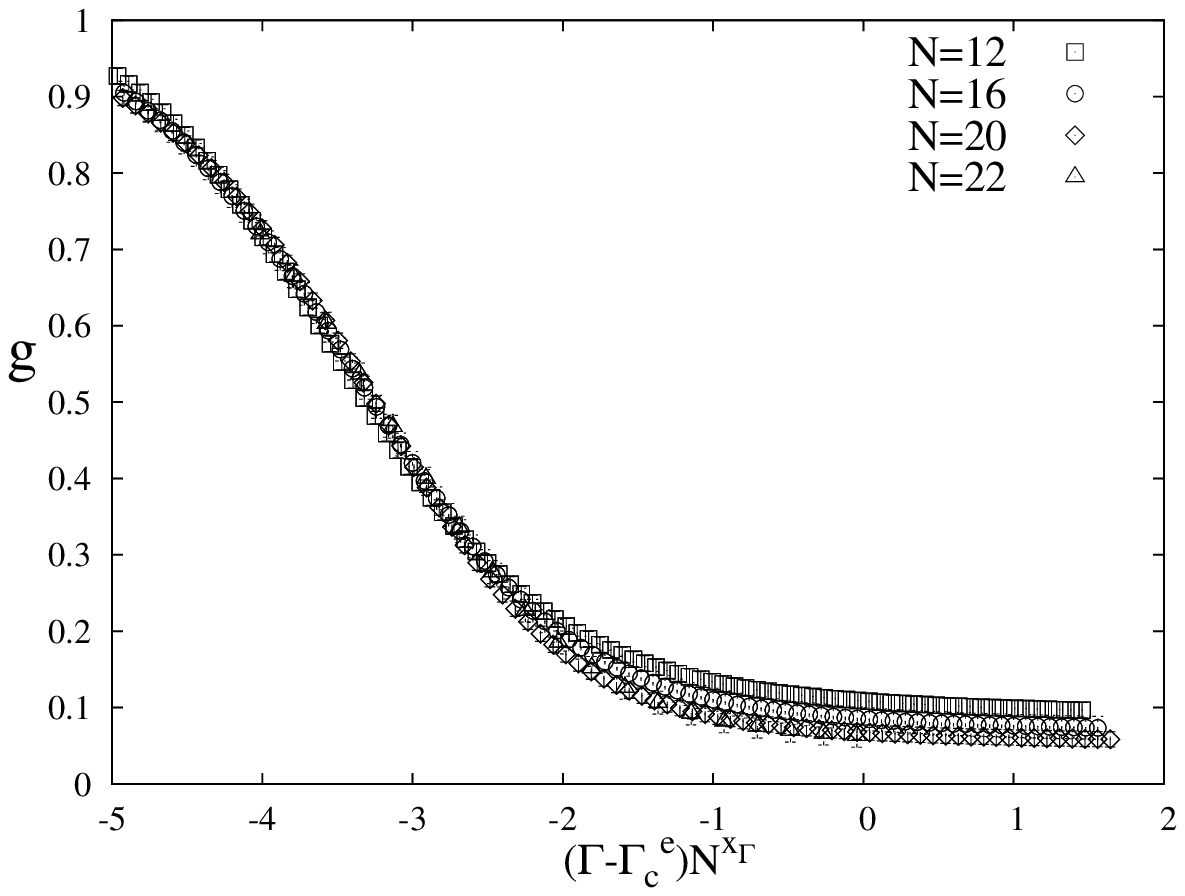}
\end{center}
\caption{(Color online) The plot (a) shows the variation of Binder cumulant $g$ as a function of $\Gamma$ for different system sizes for quantum S-K model  at $T = 0$ 
(exact diagonalization results). The larger system sizes intersect at higher values of $\Gamma$ signifying finite size effect of the system. (b) shows the Binder cumulant curves for different system sizes ($N$) collapse following the scaling fit (to Eq.~(\ref{gt}) for $M=0$) with an estimated 
$\Gamma_c^e=1.61$ and exponent $x_{\Gamma}=0.50\pm0.02$.}
\label{bc_gama}
\end{figure}
\begin{figure}[ht]
\begin{center}
\includegraphics[width=9.5cm]{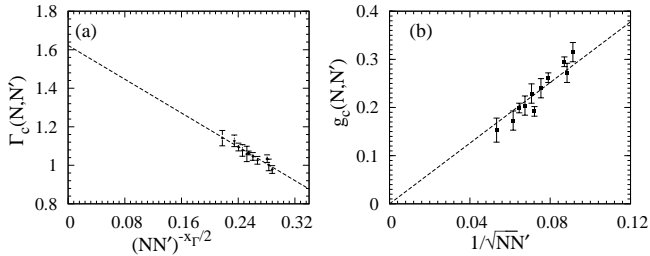}
\end{center}
\caption{The plot shows critical transverse field ($\Gamma_c(N,N')$) and critical Binder cumulant ($g_c(N,N')$) as a function of $(NN')^{-x_{\Gamma}/2}$ and $1/\sqrt{NN'}$ respectively, where $N$ and $N'$ are two system sizes, obtained from exact diagonalization: (a) The extrapolated value of $\Gamma_c$ is found to be $1.62$ and (b) $g_c$ tends to a null value for a infinite size 
system. For these two plots best fit line is also shown.}
\label{gc_bc}
\end{figure}

The variations of $g$ as a function of $\Gamma$ is shown in Fig.~\ref{bc_gama}(a) for different system sizes. To study the finite size effects, we consider a pair of two different system sizes $N$ and $N'$ and evaluate the values of $\Gamma_c(N,N')$ 
 and  $g_c(N,N')$ from the intersection of the $g$ vs. $\Gamma$ curves for these two system sizes. Accounting every possible pair, we extrapolate $\Gamma_c(N,N')$ with $(NN')^{-x_{\Gamma}/2}$ to get $\Gamma_c$ for infinite system size. In absence of an established  finite size scaling behavior of $g$, we fit its finite size variations of $g_c(N,N')$ to  $1/\sqrt{NN'}$ to evaluate $g_c$ in the thermodynamic limit.
The extrapolated value of $\Gamma_c(N,N')$ is $1.62~\pm~0.05$ in the limit of $N,N'\rightarrow\infty$ which is indicated in Fig.~\ref{gc_bc}(a).
Here the best fit value of the scaling exponent $x_{\Gamma}$ for geting the extrapolated value of $\Gamma_c(N,N')$ is $0.50\pm0.02$,
 which is consistent with that obtained from collapse of $g$ curves for different system sizes (see Fig.~\ref{bc_gama}(b)). One can also see that 
the extrapolated value of $\Gamma_c(N,N')$ is nearly equal to the estimated value $\Gamma_c^e=1.61$, which is required for getting a good collapse of Binder cumulant 
curves for different system sizes (see Fig.~\ref{gc_bc}(b)).
On the other hand $g_c$ takes nearly a zero value in the limit of $N,N'\rightarrow\infty$ (see Fig.~\ref{gc_bc}(b)), and this is consistent with our Monte Carlo results at the low temperatures.  These indicate that starting from around $T=0.35$ to $T=0$ the values of $g_c$ as well as of $x_{\Gamma}$ remain practically unchanged at its quantum fluctuation dominated value ($g_c \simeq 0, x_{\Gamma} \simeq 0.50$).

\section{Summary and Discussions }\label{summary}

\begin{figure}
\begin{center}
\includegraphics[width=6.2cm]{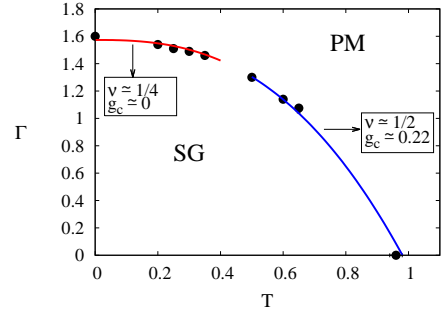}
\end{center}
\caption{(Color online) Consolidated phase diagram of the S-K spin glass model in transverse field, obtained from the Monte Carlo simulation and exact diagonalization. The statistical errors are indicated if they are more than the symbol sizes. 
Here SG and PM denote respectively the spin glass and paramagnetic phases. 
The points at $T = 0$ and $\Gamma = 0$ correspond to purely quantum and classical cases respectively. 
The obtained critical behaviors are indicated ($g_c \simeq 0$, $\nu \simeq 1/4$ for low $T$-high $\Gamma$ region and $g_c \simeq 0.22$, $\nu \simeq 1/2$ for high $T$-low $\Gamma$ region).  
The crossover point is around $T \simeq 0.45$ and $\Gamma \simeq 1.46$.}
\label{phase_diagram}
\end{figure}

In summary, we estimate the entire phase diagram (see Fig.~\ref{phase_diagram}) of the quantum S-K spin glass using Monte Carlo simulation and exact diagonalization results. The estimated phase diagram  compares well with some earlier estimates for isolated parts (Refs.~\cite{sudip-gold,sudip-lai}]). We use system sizes $N=20, 60, 180$ with moderately chosen $M$ to keep  $M/L^{z}$ constant) for  Monte Carlo simulation whereas for exact diagonalization maximum system size limits up to $N = 22$.
During the exploration of phase diagram by varying $T$ or $\Gamma$, we find that $g_c$ remains fairly constant (at value $0.22 \pm 0.02$) from classical transition point ($\Gamma = 0$, $T\simeq 1.0$) to almost $T=0.45, \Gamma = 1.33$ and assumes a very low value ($< 0.
02$) or vanishes (with Gaussian fluctuations) beyond this point and remains the same up to $\Gamma \simeq 1.62$, $T = 0$ (see also~\cite{sudip-taka}]). The scaling fits to Eq.~(\ref{gt}) give $x_T$ or $x_{\Gamma}=0.31\pm 0.02$  for high $T$ and low $\Gamma$ values, 
while   $x_{\Gamma}=0.50\pm 0.02$  for low $T$ and high $\Gamma$ values. We should mention that we performed the same Monte Carlo simulations on infinite range pure ferromagnetic system. In this case, $g_c$ and $x_{\Gamma}$ values remain almost the same for any finite temperature we considered (up to $0.1 T_c$), indicating that the crossover to quantum behavior occurs only at zero temperature, as theoretical analysis for such pure systems clearly suggests (see e.g., \cite{sudip-bkc-book}).

 We believe, these two values of $g_c$ indicate two different universality classes and our observation indicates that the universality class of classical fluctuation dominated transitions (at low $\Gamma$ and high $T$) is quite different from that for the quantum fluctuation dominated transitions (for high $\Gamma$ and $T$). Existence  of such distinct universality classes appears more reasonable when compared with the observation that the correlation length exponent $\nu$ also has two different values in these two parts of the phase boundary (having two different values of $g_c$). If we take effective dimension  $d_c=6$~\cite{sudip-Billoire} and $x_T = x_{\Gamma} = 1/3$ for entire classical fluctuation dominated transitions, then using the relation $x_{\Gamma}= x_T=1/d_c\nu$ (Eq.~(\ref{gt}), see also~\cite{sudip-david,sen_97}) we get $\nu=1/2$, which is consistent with the earlier estimate~\cite{sudip-Billoire}. Similarly for quantum fluctuation 
dominated transitions we find  $\nu = 1/4$ for  $x_{\Gamma} = 1/2$ (considering $d_c=8$~\cite{sudip-david,sudip-read}), which agrees with earlier estimate~\cite{sudip-david,sudip-read}. Such changes in the values of $g_c$ and  $\nu$ clearly indicate that, in contrast to the pure case, the crossover between classical and quantum fluctuation dominated critical behaviors for the transverse Ising S-K model occurs at a non-vanishing temperature.

Due to random and competing spin-spin interactions, the free energy landscape of S-K spin glass is highly rugged. Such uneven free energy landscape contains high ($O(N)$) free energy barriers, which separate several local free energy minima. For low $T$, unlike in the pure case (where the landscape is inclined smoothly towards the minima), the thermal fluctuations become ineffective in helping such systems  to cross tall barriers to reach the paramagnetic state by flipping finite fractions of $N$ spins. On the other hand due to the presence of high $\Gamma$, tunneling through such tall but narrow barriers becomes highly probable~\cite{sudip-ray,sudip-Heim}. Quantum fluctuations therefore induce the phase transition and determine the transition behavior. Such effectiveness of quantum (over thermal) fluctuations at low $T$ in such frustrated systems might therefore be responsible for a classical-quantum crossover at a finite (but low) temperature (and large transverse field) in the quantum S-
K 
model. In fact, our study establishes that the critical value of the Binder cumulant (with associated scaling exponents) and its crossover behavior gives a quantitative measure of the relative importance of classical versus quantum fluctuations in determining the nature of the phases in such frustrated systems.

\acknowledgements
We are grateful to Purusattam Ray, Parongama Sen and Sabyasachi Nag for their comments and suggestions.



\begin{thebibliography}{100}
\bibitem{sudip-sk}
 K. Binder and A.P. Young, Rev. Mod. Phys. \textbf{58}, 801 (1986).
 \bibitem{sudip-bkc-book}
 S. Suzuki, J.-i. Inoue, and  B. K. Chakrabarti, \textit{Quantum Ising Phases $\&$ Transitions in Transverse Ising Models}, Springer, Heidelberg (2013); 
 A. Dutta, G. Aeppli, B. K. Chakrabarti, U. Divakaran, T. Rosenbaum and D. Sen, \textit{Quantum Phase Transitions in Transverse Field Models}, 
 Cambridge Univ. Press, Delhi (2015).
 \bibitem{sudip-yamamoto}
 T. Yamamoto and H. Ishii, J. Phys. C. \textbf{20}, 35 (1987).
 \bibitem{sudip-usadel}
 K. Usadel and B. Schmitz, Solid State Commun. \textbf{64}, 6 (1987).
 \bibitem{sudip-kopec}
 T. K. Kopec,  J. Phys. C. \textbf{21}, 2 (1988).
 \bibitem{sudip-gold}
 Y. Y. Goldschmidt and P. Y. Lai, Phys. Rev. Lett. \textbf{64}, 2467 (1990).
 \bibitem{sudip-lai}
 P-Y. Lai and Y. Y. Goldschmidt,  Europhys. Lett. \textbf{13}, 289 (1990).

 
 \bibitem{sudip-binder}
 K. Binder and D. Heermann, \textit{Monte Carlo Simulation in Statistical Physics}, Springer, Heidelberg (2010).
 \bibitem{sudip-guo}
M. Guo, R. N. Bhatt, and D. A. Huse, Phys. Rev. Lett. \textbf{72}, 4137 (1990).


\bibitem{sudip-alvarez}
J. V. Alvarez and F. Ritort, J. Phys. A: Math. Gen. \textbf{29}, 7355 (1996).

\bibitem{sudip-Billoire}
 A. Billoire and I. A. Campbell, Phys. Rev. B. \textbf{84}, 054442 (2011).
 
 \bibitem{sudip-david}
 D. Lancaster and F. Ritort, J. Phys. A: Math. Gen, \textbf{30}, L41 (1997).
 
\bibitem{sudip-read}
N. Read, S. Sachdev  and J. Ye, Phys. Rev. B. \textbf{52}, 384 (1995).
\bibitem{lanczos_52}
C. Lanczos, J. Res. Nat. Bur. Stand. \textbf{49}, 33 (1952).

\bibitem{sudip-taka}
K. Takahashi, Phys. Rev. B. \textbf{76}, 184422 (2007).

\bibitem{sen_97}
 P. Sen, P. Ray, and B. K. Chakrabarti, arXiv:cond-mat/9705297 (1997).


\bibitem{sudip-ray}
P. Ray, B. K. Chakrabarti  and A. Chakrabarti, Phys. Rev. B. \textbf{39}, 11828 (1989).
 
 \bibitem{sudip-Heim}
B. Heim, T. F. Ronnow, S. V. Isakov, and M. Troyer,   Science \textbf{348}, 215 (2015).
 
 \end{thebibliography}
\end{document}